\documentstyle[12pt]{article}
\begin{document}
\newcommand{\ket}[1] {\mbox{$ \vert #1 \rangle $}}
\newcommand{\bra}[1] {\mbox{$ \langle #1 \vert $}}
\newcommand{\bkn}[1] {\mbox{$ < #1 > $}}
\newcommand{\bk}[1] {\mbox{$ \langle #1 \rangle $}}
\newcommand{\scal}[2]{\mbox{$ \langle #1 \vert #2  \rangle $}}
\newcommand{\expect}[3] {\mbox{$ \bra{#1} #2 \ket{#3} $}}
\newcommand{\ki}{\mbox{$ \ket{\psi_i} $}}
\newcommand{\bi}{\mbox{$ \bra{\psi_i} $}}
\newcommand{\p} \prime
\newcommand{\e} \epsilon
\newcommand{\la} \lambda
\newcommand{\om} \omega   \newcommand{\Om} \Omega
\newcommand{\cc}{\mbox{$\cal C $}}
\newcommand{\w} {\hbox{ weak }}
\newcommand{\al} \alpha
\newcommand{\bt} \beta

\newcommand{\be} {\begin{equation}}
\newcommand{\ee} {\end{equation}}
\newcommand{\ba} {\begin{eqnarray}}
\newcommand{\ea} {\end{eqnarray}}

\def\lrD{\mathrel{{\cal D}\kern-1.em\raise1.75ex\hbox{$\leftrightarrow$}}}

\def\lr #1{\mathrel{#1\kern-1.25em\raise1.75ex\hbox{$\leftrightarrow$}}}

\overfullrule=0pt \def\sqr#1#2{{\vcenter{\vbox{\hrule height.#2pt
          \hbox{\vrule width.#2pt height#1pt \kern#1pt
           \vrule width.#2pt}
           \hrule height.#2pt}}}}
\def\square{\mathchoice\sqr68\sqr68\sqr{4.2}6\sqr{3}6} 
\def\lrpartial{\mathrel
{\partial\kern-.75em\raise1.75ex\hbox{$\leftrightarrow$}}}

\begin{flushright}
LPTENS 96/45\\
october 1996
\end{flushright}
\vskip 1. truecm
\vskip 1. truecm
\centerline{\LARGE\bf{Time dependent Green functions}}
\vskip 2 truemm
\centerline{\LARGE\bf{in Quantum Cosmology}}
\vskip 1. truecm
\vskip 1. truecm

\centerline{{\bf R. Parentani}\footnote{
present address: Laboratoire de Math\'ematique et Physique Th\'eorique,
Facult\' e des Sciences,
\newline 
${ \quad\quad\quad\quad\quad\quad
\quad\quad\quad } $  Universit\'e de Tours, 37200 Tours, France}}

\vskip 5 truemm
\centerline{Laboratoire de Physique Th\'eorique de l'\' Ecole
Normale Sup\'erieure\footnote{Unit\'e propre de recherche du C.N.R.S.
associ\'ee \`a l'ENS et \` a l'Universit\'e de Paris Sud.}}
\centerline{24 rue Lhomond,
75.231 Paris CEDEX 05, France.}
\centerline{
e-mail: parenta@celfi.phys.univ-tours.fr}
\vskip 5 truemm
\vskip 1.5 truecm

\vskip 1.5 truecm
{\bf Abstract }
The aim of this article is twofold.
First we examine from a new angle 
the question of the recovery of time 
in 
quantum cosmology.
We construct Green functions for matter fields 
from the solutions
of the Wheeler De Witt equation. For simplicity we work in 
a mini-superspace context. By evaluating these Green functions
in a first order development of the energy
{\it increment} induced by matrix elements of field operators,
we show that the background geometry is the solution 
of Einstein equations driven by the
mean matter energy and that it is this  background 
which determines the time lapses separating
the field operators.
Then, by studying higher order corrections, we clarify the nature 
of the small dimensionless parameters which 
guarantee the validity of the approximations used. 
In this respect, we show that the formal expansion in 
the inverse Planck mass which is sometime
presented as the ``standard procedure'' is illegitimate.
Secondly, by the present analysis of Green functions,
we prepare the  
 study of quantum matter transitions
in quantum cosmology.
In a next article, we show that 
the time parametrization of transition amplitudes
appears for the same reasons
that it appeared in this article. This proves that the background is
dynamically determined by the transition under examination.

\vfill \newpage

\section{Introduction}

In General Relativity, the invariance under reparametrizations
in time and in space leads to a set of local constraints. These
impose that the total  four-momentum of matter + gravity
vanishes identically. Therefore, in contrast to the usual
classical and quantum mechanics, there is no preferred 
reference frame to specify when and where events occur.
This absence of external reference frame can also be 
conceived as arising from the dynamical character of space-time.
Hence, since space-time is 
coupled to all matter transitions, it is 
coupled to the particular event we want to localize. 
Therefore,  
in a weak perturbation limit, 
the matter-gravity correlations resulting from the
constraints should deliver the historical description of events
performed in the background geometry.
Then two 
questions arise:

How to extract from the total wave function of the universe the 
properties of
the background geometry in which time lapses are defined ?

Under what circumstances is the extraction of a non-dynamical
background a legitimate approximation ?

Both questions have received 
attention
and we refer to \cite{isham}\cite{hartle}\cite{ontime} for
 reviews in which  various attempts are compared.
We recall that it has been shown in 
\cite{rub}\cite{banks}\cite{HalH} how, 
starting from the solutions of the Wheeler-De Witt equation,
the semi-classical character of gravity 
leads to the Schroedinger equation for matter fields.
This treatment has been pursued in 
\cite{brout}\cite{Hal}\cite{vil}\cite{BV}\cite{kiefer1} 
and both higher
order corrections and the criteria for having reliable
approximations have been discussed.
However there is no consensus on the manner to
identify the
background.
In refs. \cite{banks}\cite{kiefer1}\cite{kiefer2}\cite{ortiz},
the development in a power series
of $1/G$ (Newton's constant)
leads inevitably to backgrounds which are
 {\it empty} solutions of Einstein equations.
Instead, in refs. \cite{HalH}\cite{brout}\cite{BV}, it was proposed that
the background solution must contain the {\it mean} matter energy
as a source, in a manner similar to what happens in a 
Born-Oppenheimer treatment: in the 
adiabatic approximation, the protons feel the {\it mean} Coulomb potential 
delivered by the electrons.
 
In this paper, we settle this debate and
prove, in conformity with  \cite{HalH}\cite{vil}\cite{BV}, 
that the background solution 
is the Einstein solution driven by the mean matter content.
To reach this conclusion we proceed as follows.
Instead of focusing on the wave function
of the universe itself, in which matter and gravity are
treated on a same footing, we center our attention
on matter matrix elements
evaluated at some given gravitational 
configuration. In particular, we analyze matter
Green functions wherein the two insertions of the
field operator are performed at two different radii
of the universe.
This un-symmetrical treatment in which 
a restricted number of matter quanta are involved
determines unambiguously
what is the
background geometry and 
what are the developments which 
lead to the localization of 
quantum matter effects.
%
%
The small parameter controlling these developments 
is  the 
{\it relative change} of matter energy induced by the
insertion of the field operators appearing in the
Green function itself.

This is as it should be: given a certain matrix element, 
the validity of the background field approximation 
puts constraints on the {\it sources} of the background field, 
see Chapter 14 
in \cite{jr}.
Therefore the Planck mass has nothing to do with its validity 
nor does it govern its corrections.
The 
corrections to the background field approximation 
are of two kinds.
One has the corrections to the WKB approximation for the
propagation of gravity, see 
\cite{kiefer1},
 and the nonlinear dependence of the geometry in 
the energy change. 
At this point, it is relevant to notice the
close analogy between the present development leading 
to the concept of {\it time} 
and the procedure by which a big isolated system furnishes
a {\it temperature} to a little system contained in it. 
Both equilibrium concepts, i.e. time and inverse temperature, 
are obtained through a {\it first order} change in the energy
and determined by the properties of the big system. Moreover, the
second order corrections, which control the validity of these equilibrium 
notions, have the same origin: the big system is
finite and has therefore a finite specific heat. 

Very important also is that 
only a small part of the corrections to 
the WKB approximation affects the value of
the Green functions.
Indeed, only that part which
 is sensitive to the matter energy change modifies
the Green functions. This has two important consequences.
First, for matter dominated universes, the modifications
of the Green functions due to these  WKB corrections are
irrelevant since they are one higher order than modifications induced 
by the non-linearities. Secondly,
the validity of the WKB approximation is much
less restrictive that the one applied on the full
wave function of the universe. 
 
Finally, we clarify the role of the spread of the matter energy
in the universe. In former works, see e.g. \cite{BV}\cite{ortiz}, 
the problem of the recovery of time 
was considered together with the validity of representing gravity by
a {\it single} background. 
In this work, we emphasize that these two problems 
should be considered separately.
Indeed we show that {\it each} component 
of the wave function defines its own time lapse. 
Therefore large spreads are 
perfectly acceptable. Simply the notion of a 
{\it single} background 
defining a {\it single} time
valid for all configurations looses sense.
But this has no physical consequences, since
remote configurations do not interfere.


In the next paper\cite{wdwpt},
we define and evaluate the amplitudes
of matter transitions in the framework of quantum cosmology.
Upon evaluating these amplitudes, we shall see 
that a time parametrization 
is provided for the same reasons that time appeared in
our Green functions, {\it i.e.} through a first order change 
of the energy change induced by the transition itself. We emphasize
this point:
Time appears 
because there is a transition,
more precisely, time appears because there is an 
interaction hamiltonian which determines the
probability amplitude to have a transition.
This legitimizes dynamically the analysis of the present paper
which is purely kinematical.

In this paper we proceed as follows.
We consider two different matter fields,
a massive one and a massless one
conformally coupled.
This allows us 

a.  to reveal how, by using gauge independent quantities,
 proper time lapses characterize Green functions of the massive field
whereas it is the conformal time which appears in the massless 
field's functions;

b.  to show how the appearance of these time lapses
relies on a {\it double} development:
a WKB approximation for 
the kernel of gravity together with a first order development
in the change of energy induced by the field
operator involved in the Green function;

c. to show how the spectator field, 
not directly involved in the Green function,
determines indirectly this function 
through its energy. Indeed, its energy contributes
to the determination of the background
geometry in terms of which time lapses are, in turn,
defined; and


d. to analyze the corrections to these approximations
and the physical conditions under which these corrections can be
correctly discarded.

All of these developments are done
in the mini-superspace context, i.e.
both the field configurations and the field operators are homogeneous
and carry zero momentum.
At the end of the article, we consider an extension of the formalism
in which the momenta no longer vanish.


\section{Why study Green functions ?}

In this section
we compare the time dependence
of the Green functions of an harmonic oscillator
to the time dependence of its wave function.
The technical reason for comparing these well known features 
is to make the reader aware of the different roles 
that the total matter
energy and the energy change induced by the insertions of the fields operators
will play upon considering, in Section 4,
Green functions from the solutions of the
Wheeler De Witt (WDW) equation.
In that Section, we shall see that the total matter
energy determines the background solution whereas
the energy change characterizes the phase of the Green function.

The physical justification for considering Green functions is the
usual one: 
their study prepares the analysis of transitions amplitudes
among matter constituents in quantum field theory. To 
study these  transitions amplitudes in the framework of
quantum cosmology is precisely the subject of \cite{wdwpt}.

In non-relativistic quantum mechanics,
in the Schroedinger picture, the time dependence
of the state of the system is governed
by the following equation
\ba
\ket{\Psi(t)}= e^{-i H (t-t_0)/ \hbar } \ket{\Psi_0}
\label{one}
\ea
where $\ket{\Psi_0}$ is the wave function specified at $t=t_0$.

When the system is in an eigenstate of the energy,
this equation reduces to 
\ba
\ket{\Psi(t)}= e^{-i E (t-t_0)/ \hbar  } \ket{\Psi_0}
\label{two}
\ea
where $E$ is the {\it {total}} energy of the system.
For instance, for a harmonic oscillator in the n-th 
level, one has $E=E_n = \hbar \om \ (n+1/2)$.

We now consider the time dependence of Green functions.
This is usually done in the Heisenberg picture upon dealing
with the time dependent Heisenberg operators
$q(t) =e^{i H (t-t_0)/ \hbar  }  q  e^{-i H (t-t_0)/ \hbar  } $. 
 For instance, for an
harmonic oscillator, whose state at $t=t_0$ is $\ket{\Psi_0}$,
the Green function is given by
\be
G_q(t_2, t_1) = \expect{\Psi_0}{  
q(t_2) q(t_1) 
 }{\Psi_0}
\label{three}
\ee
When $\ket{\Psi_0}$ is the eigenstate $\ket{n}$ of energy $E_n$,
one finds
\be
G^{n}_q(t_2, t_1) = {1 \over 2 \om} 
\left( 
n \ e^{i \om (t_2-t_1)}  + (n+1) \ e^{-i \om (t_2-t_1)}
\right)
\label{four}
\ee
The differences between state propagation in eq. (\ref{two})
and Green function time dependence are manifest
upon computing eq. (\ref{four}) in the Schroedinger picture:
\ba
G^{n}_q(t_2, t_1) &=& \expect{n} {e^{i H (t_0-t_2)/ \hbar  } \ q \ 
e^{-i H (t_2-t_1)/ \hbar  } \ q \   e^{-i H (t_1-t_0)/ \hbar  }} {n}
\nonumber\\
&=& \expect{n} {q \ 
e^{-i H (t_2-t_1)/ \hbar  } \ q \   e^{-i H (t_1-t_2)/ \hbar  }} {n}
\nonumber\\
&=& 
\left[ \expect{n}{{  a^\dagger  a \over 2 \om }}{n}\
e^{- i (E_n -\hbar  \om) (t_2-t_1)/ \hbar  } 
+ 
\expect{n} { {a  a^\dagger   \over 2 \om } }{n} \ 
e^{- i (E_n + \hbar \om) (t_2-t_1)/ \hbar  } \right]
 e^{i E_n (t_1-t_2)/ \hbar  } 
\nonumber\\
&=& 
{ n \over 2 \om} \ e^{i \om (t_2-t_1)}  + {n+1 \over 2 \om} \ e^{-i \om (t_2-t_1)}
\label{fur}
\ea
where we have introduced the (Schroedinger) time independent
expectation value $\expect{n}{a  a^\dagger  }{n}=n+1$.
The total energy $E_n$ drops out the phase of the Green function.
Only $\pm \om$, the energy changes induced by the $q$ 
operators, survive.
 This will
be extremely important in the quantum cosmology. In that context, 
the Green functions are delivered in 
the Schroedinger form, as in the third line of eq. (\ref{fur}). 
Furthermore, the two independent matter energy scales $ \hbar \om$,
 the (microscopic) {\it change}  in energy induced by the
operator $q$, and 
$E_n = n \hbar  \om$, the (macroscopic) energy in the universe, 
play completely different roles\cite{vil}.
Indeed, it is by a development in $1/n$ 
that the background solution characterized by 
the total energy is isolated. Furthermore, higher order terms
in $1/n$ determine the corrections induced by the fact that 
geometry is dynamical and can only be approximated 
by a single passive background.

The other difference is that 
the Green function is independent of $t_0$, i.e. of the
initial phase of the wave function and depends only in the
lapse between $t_2$ and $t_1$. 
This local character will also play an important role in the
WDW context since only local corrections will
enter into the nonlinear response of gravity.

However it should be pointed out
that upon dealing with 
a superposition,
$\ket{\Psi_0} = \Sigma_n c_n \ket{n}$,
a dependence in $t_0$ and $t_2+ t_1$ does appear.
Indeed one obtains
\ba
G^{c_n}_q(t_2, t_1) &=& 
\Sigma_n |c_n|^2 {1 \over 2 \om}  \left( n \ e^{i \om (t_2-t_1)}
+ (n+1) \ e^{- i \om (t_2-t_1)} \right)
\nonumber\\
&& + 
\Sigma_n \mbox{2 Re}\left[ {1 \over 2 \om}  c_n c_{n-2}^* \sqrt{ n (n-1)} 
\ e^{- i \om ( t_2 + t_1 - 2 t_0)}
\right]
\label{five}
\ea
The dependence on $t_0$ and $t_2+ t_1$ emerges from the
{\it interfering} terms relating $\ket{n}$ to $\ket{n \pm 2}$.
Therefore, when the system
is described by a density matix which is diagonal in 
energy, i.e. $\hat \rho = \Sigma_n \rho_n \ket{n} \bra{n}$,
the interfering terms do not exist and one has
\be
G^\rho_q (t_2, t_1) = Tr [ \ \hat \rho  \ q(t_2) q(t_1)  ]
 = \Sigma_n \rho_n  {1 \over 2 \om}\left( 
n \ e^{i \om (t_2-t_1)}  + (n+1) \ e^{-i \om (t_2-t_1)}
 \right)
\label{six}
\ee
Because of this simplicity, in Section 4,
 we shall first consider 
Green function arising from diagonal
density matrices and only then discuss 
the new aspects brought in by the 
interfering terms.

\section{The cosmological model}

We recall the main 
features of the dynamics
in mini-superspace, see {\it e.g.} \cite{isham}\cite{hartle}. We put
special emphasis on
the relationship between the matter energy content of a universe
and lapses of proper time evaluated in that universe. 
The reason for this emphasis is that the time parametrization
reappears in quantum cosmology through this 
relationship.

To make this as clear as possible,
we choose the matter content in a such a way that
its state is characterized by constants of motion.
This allows to write the Hamilton-Jacobi action as a sum of actions
and the wave function as a sum of products.
Because of the simplicity of this model,
the modifications introduced by the
quantum character of gravity will be easily
identified and evaluated. 
More elaborate situations will be consider afterwards.
In Appendix A, we leave the mini-superspace restriction and 
 in \cite{wdwpt}, we introduce interactions which cause
the ``constants of motion'' to evolve.

We consider a cosmological
model with both 
massive and massless matter fields.
For the massive field, we impose that the mass $M$ be large 
enough that its Compton wave length satisfies
\be
{\hbar \over M } << {a \over {\dot a}}
\label{M1}
\ee
where $\dot a = da /dt$ with $t$ being the proper time 
of the universe, i.e. when the lapse satisfies
$N=1$. We recall that, in mini-superspace, the metric element reads
\be
ds^2 = - N^2(\xi) d\xi^2 + a^2(\xi) d^2 \Omega_3
\label{M2}
\ee
where $N(\xi)$ is the lapse and
where $d^2 \Omega_3$ is the constant line element of the
homogeneous three surfaces.

When eq. (\ref{M1}) is satisfied, there is no pair production
of massive quanta and 
the action of an homogeneous massive scalar field  $\tilde \psi(t) $
is correctly approximated by (from now on we put $\hbar = 1$)
\be
S_M = {1 \over 2 } \int dt  \left( \dot \psi ^2 - M^2 \psi ^2 \right)
\label{M3}
\ee
where the field $\psi(t)$ is related to the original scalar 
$\tilde \psi(t) $ by $\psi(t) = a^{3/2}(t) \tilde \psi(t) $.
Indeed one verifies that the differences between the
original field equation and the harmonic (adiabatic)
equation for $\psi(t)$
scale like $(\dot a / a M)^2 << 1$. Non adiabatic corrections, which
govern pair creation amplitudes, will be considered in \cite{wdwpc}.

For the massless conformally coupled field, see \cite{BD}, its action is
\be
S_\gamma =  {1 \over 2 } \int d\eta \left( 
(\partial_\eta \phi)^2 - k^2  \phi^2  \right) =
{1 \over 2 } \int { dt \over a } \left(   a^2
 \dot \phi^2 - k^2  \phi^2  \right) 
\label{M4}
\ee
where $d\eta = dt /a(t)$ defines the conformal time. 
It that gauge, $N(\eta)=a(\eta)$. The field $\phi$ is related to the
original massless field $\tilde \phi$ by $\phi = a  \tilde \phi$.

The total energy of matter (when $N=1$) is given by
\be
H_{m}(a) = M |A_M |^2 + { k \over a }  |A_\gamma |^2 
\label{M5}
\ee
where $A_M , A_\gamma $ are the classical 
amplitudes of the
$\psi$ field and the $\phi$ field whose norms are {\it conserved},
i.e. are $a$ independent.

The action for gravity is given by the Einstein lagrangian.
When restricted to the conformal mode in the proper time gauge 
 is reduces to
\be
S_G = {1 \over 2G} \int dt \left( - a \ \dot a^2 
- \kappa a - \Lambda a^3 \right)
\label{M5'}
\ee
where $\kappa$ is equal to $0, \pm$ for flat, open or closed three 
surfaces,
where $\Lambda$ is a cosmological constant and where $G$ is 
Newton's constant.
Notice the minus sign of the kinetic term of $a$. This is a 
common feature
of reparametrization invariant theories.
In the present case, the total hamiltonian of gravity plus matter
 is constrained to vanish in order to guarantee the
 invariance under time reparametrizations. In terms of 
the momentum of gravity, $\pi_a$, this constraint is given by
\be
H_G + H_{m}= \left({
-G^2 \pi_a^2 + \kappa a^2 + \Lambda a^4 
\over 2Ga }  \right) + 
\left(M |A_M |^2 + { k \over a }  |A_\gamma |^2 \right)
=0
\label{M6}
\ee
where $\pi_a$ is related to $a$ and to
the proper time $t$ by
\be
\pi_a = -{ a \dot a \over G}
\label{M7}
\ee

The use of the momentum $\pi_a$ allows to write 
the constraint without any reference to the lapse $N$,
i.e. in a gauge independent manner. Thus, for a given
matter content, in our case with $A_M , A_\gamma $ specified,
$\pi$ is determined (up to a sign) by the constraint eq. (\ref{M6})
\be
\pi ( a, E_M  , E_\gamma ) = - G^{-1} \sqrt{ 
\kappa a^2 + \Lambda a^4 + 2 Ga 
\left(E_M + { E_\gamma \over a } \right) }
\label{M8}
\ee
where $E_M = M |A_M|^2$ is the energy of the massive field
and $E_\gamma = k |A_\gamma|^2$ the ``conformal''
energy of the massless field.
The sign ambiguity is fixed by restricting ourselves to 
expanding universes, i.e. $\dot a > 0 $.
Therefore,  instead of determining $\pi$ in terms of a given $a(t)$, 
eq. (\ref{M7}) should be considered as determining
the proper time $t$ in terms of $\pi$ and $a$
having 
specified the matter content.
Indeed one has
\ba
\Delta t( a_2, a_1 ;  E_M,  E_\gamma )
&=& - \int_{a_1}^{a_2} da { a \over G \ \pi( a,  E_M, E_\gamma ) } 
\nonumber\\
&=& - \partial_{E_M} \int_{a_1}^{a_2} da \ \pi( a,  E_M, E_\gamma ) 
\label{M9}
\ea
Similarly, the lapse of conformal time 
$\Delta \eta$ 
is determined by 
\ba
\Delta \eta( a_2, a_1 ;  E_M,  E_\gamma )
&=& - \int_{a_1}^{a_2} da { 1  \over G \ \pi( a,  E_M, E_\gamma ) } 
\nonumber\\
&=& - \partial_{E_\gamma} \int_{a_1}^{a_2} da \  \pi( a,  E_M, E_\gamma ) 
\label{M10}
\ea

In the second lines of eqs. (\ref{M9}, \ref{M10}), we have expressed 
the fact that
the proper time $t$ is, {\it by definition},
 the conjugate variable of the energy $E_M$ and
 the conformal time $\eta$ is the conjugate variable of the
conformal energy $E_\gamma$. In both cases, it is the gravitational 
part of the action,
i.e. $S_G=\int da \ \pi_a $,
see below eq. (\ref{M12'}), which is used to define the conjugate variable.
By using the angle-action terminology, we can say that the proper
time is the ``angle'' conjugated to the ``action'' $E_M$, see e.g.
\cite{mtw} for a brief and clear presentation of these concepts.
When using the solutions of the WDW equation to evaluate Green 
functions in quantum cosmology, we shall
see that it is through these equations
that time lapses will be defined and determined, answering therefore
point a. of the Introduction.
Moreover,
these Hamilton-Jacobi equations make manifest the analogy
with the inverse temperature which is defined in a 
microcanonical ensemble
as the derivative of the entropy with respect to the energy, 
see eq. (\ref{betab}).
 
Since $A_M$ and $ A_\gamma $ are conserved, the Hamilton-Jacobi action
is a {\it sum} of three actions
\be
S(a, \psi, \phi ) = S_G(a ; E_M ,  E_\gamma ) + S^{o.h.}_M ( \psi  ; A_M)
+ S^{o.h.}_\gamma ( \phi ; A_\gamma )
\label{M11}
\ee
where the matter actions are the actions for  harmonic oscillators
of amplitude $A$ and respective frequency $M$ and $k$.
The gravitational part of the action satisfies
\be
-G^2  ( \partial_a S_G (a) )^2
 + \kappa a^2 + \Lambda a^4 + 2 Ga 
\left(  E_M
+ { E_\gamma \over a } 
\right)
=0
\label{M12}
\ee
Since $S_G$ depends on a single variable only,
 it is entirely determined by the classical momentum,
eq. (\ref{M8}), by
\be
S_G(a ; E_M,  E_\gamma ) = 
\int_{a_1}^{a_2} da \ \pi( a, E_M,  E_\gamma  ) 
\label{M12'}
\ee

Upon quantizing the system, the conservation of the energies
$E_M$ and $ E_\gamma$ implies that
the wave function factorizes 
into a sum of {\it products} of waves
\be
\Psi ( a, \psi, \phi ) = \Sigma_{n_M, n_\gamma} c_{n_M, n_\gamma}
\Psi ( a; n_M, n_\gamma ) \scal{\psi}{{n_M}} \scal{\phi}{{n_\gamma}}
\label{M13}
\ee
The matter states $\ket{n_M}$ and  $\ket{n_\gamma}$ are states
of harmonic oscillators characterized by the (conserved)
number of excitations
$n_M$ and $n_\gamma$. They keep their usual 
normalization and interpretation
as providing the quantum state of the oscillators, i.e.
the interpretation that one uses when working with quantum
matter fields in a classical geometry\cite{BD}. 
This remains true when the states are adiabatically distorted
by gravity, see Appendix A.

Then, in order for the
coefficients $c_{n_M, n_\gamma}$ to keep also their interpretation
as the probability amplitude to find the matter state in the 
$n_M, n_\gamma$ sector, a normalization of the
 gravitational part of the wave function, $\Psi ( a; n_M, n_\gamma )$
must be chosen.
This wave satisfies the WDW equation at fixed $n_M$ and $n_\gamma$
\be
\left[
G^2 
\partial^2 _a  
+ \kappa a^2 + \Lambda a^4 + 2 Ga E_{m}(n_M , n_\gamma, a ) 
\right]
 \Psi ( a; n_M, n_\gamma ) =0
\label{M14}
\ee
where $E_{m}(n_M , n_\gamma, a ) = M (n_M + 1/2) 
+  k (n_\gamma + 1/2)/a $ is the eigenvalue of the matter hamiltonian,
eq. (\ref{M5}).
Our aim is not to provide a new interpretation 
to $\Psi ( a; n_M, n_\gamma )$ 
nor to discuss the ``normal ordering''
problem, i.e. to define the operator $\partial^2 _a  $.
Instead our aim is to determine (i) how to normalize $\Psi$ in order
to be able to interpret the $c_{n_M, n_\gamma}$ as 
amplitudes of probability (ii) under which conditions
its properties 
lead to a localization of quantum matter 
transitions similar to the one used when one works in
a background geometry and (iii) what are the origin
and the meaning of the corrections to this approximate description.
We refer to \cite{isham}\cite{hartle}\cite{vil} for discussions concerning
possible interpretations of the solutions of the WDW equation.

We shall prove that
the necessary condition to interpret the $c_{n_M, n_\gamma}$ as 
amplitudes of probability is that the waves $\Psi ( a; n_M, n_\gamma )$
have positive unit (conserved) wronskian, i.e.
\be
\Psi^* ( a; n_M, n_\gamma )\ 
i\!\!\lr{\partial_{a}} \Psi ( a; n_M, n_\gamma ) = 1
\label{norme1}
\ee
for all $n_M, n_\gamma $.
When considering the solutions of the Klein-Gordon equation,
this condition with $t$ replacing $a$ means,
 in the absence of pair production, i.e. when the WKB approximation 
is valid, that one deals with a single particle. 
Similarly, in the present case, it means that one works with
a single universe.

In order to prove that unit wronskian allows that interpretation,
we shall not work with 
the functions $\Psi ( a; n_M, n_\gamma )$ directly. Instead we shall
work with a kernel of the form 
$K= \Psi^* \Psi  $. The reason for this choice is that
matter matrix elements are governed by $K$ and not 
$\Psi ( a; n_M, n_\gamma )$. Indeed, both quantum matter
field theory and the
gravitational corrections induced by the dynamical character of 
$a$ will be delivered by the properties of $K$ and not those of $\Psi$.
Furthermore, the choice of unit wronskian is elegantly 
introduced and justified by gauge invariance
when working with $K$.

Indeed, in strict analogy 
with the Feynman boundary condition for relativistic
particle in the Schwinger 
formalism\cite{schwi},
one introduces a ``fifth'' time, here it corresponds to a constant
lapse $N$, and one writes the 
Schroedinger equation (first order) for the kernel 
$\tilde K(a_2,  a_1; n_M, n_\gamma ; N)$
to propagate from $a_2$ to $a_1$ in a time $N$
\be
i\partial_N - \left[ G^2 \partial^2 _a  
+ \kappa a^2 + \Lambda a^4 + 2 Ga 
E_{m}(n_M , n_\gamma, a ) 
 \right] 
\tilde K(a_2,  a_1; n_M, n_\gamma ; N) =  \delta (N) \delta (a _2 - a_1) 
\label{scwig}
\ee
This $N$ dependent kernel satisfies the usual boundary condition 
\be
\lim_{N \to 0^+}  \tilde K(a_2,  a_1; n_M, n_\gamma ; N) = \delta (a _2 - a_1) 
\label{bcond}
\ee
Then 
one chooses the 
integration contour of $N$. In strict analogy to the Feynman boundary 
condition,
we integrate over positive $N$ only and define
\be
K (a_2, a_1 ; n_M, n_\gamma )  = G^2 \int_0^\infty \!dN \ 
\tilde K(a_2,  a_1; n_M, n_\gamma ; N)
\label{M15}
\ee
where the factor of $G^2$ has been introduced for further convenience.

We point out two properties of this kernel
which arise from the second order character of the WDW equation
and the 
 integration over positive $N$ only.
First, its normalization automatically 
leads to
positivity unit wronskian, in the following sense
\be
\lim_{a_2 - a_1 \to 0^+} 
\left( i\partial_{a_2} - i\partial_{a_1} \right)
K (a_2, a_1 ; n_M, n_\gamma ) = 1
\label{M17}
\ee
Secondly, the ``convolution'' of two kernels satisfies 
\be
K (a_3, a_1 ; n_M, n_\gamma ) = K (a_3, a_2 ; n_M, n_\gamma ) \
i\!\! \lr{\partial_{a_2}} 
K (a_2, a_1 ; n_M, n_\gamma ) 
\label{M18}
\ee
when $a_3 > a_2 > a_1$.
This relation coincides with the convolution in time 
of the Feynman Green function of a relativistic particle, and 
not with the convolution of non-relativistic kernels.
The reason for the discrepancy is the additional integration over
the ``fifth'' time $N$ which projects onto zero energy solutions.
We shall see in the next Section how these two
properties guarantee that 
quantum matter field theory in a geometrical background will 
re-appear in a well defined scheme of approximations.
In this respect, there is no necessity to provide an
interpretation
 to $K(a_2, a_1)$. 
To determine how to normalize and to use it is sufficient.

Finally we rewrite this kernel introducing the
density matrix of the matter. The reason is to unify the
notations for matter Green function together with propagation of gravity.
In the case of eq. (\ref{M15}), the density matrices are
trivial: $\hat \rho_M = \ket{n_M} \bra{n_M}$ and 
$\hat 
\rho_\gamma = \ket{n_\gamma} \bra{n_\gamma}$, but the following 
expression is valid for any density matrix.
We express $K( a_2, a_1 ; \rho_M, \rho_\gamma )$ as
\be
K( a_2, a_1 ; \rho_M, \rho_\gamma ) = \tilde Tr [ \ \hat \rho_M 
 \hat \rho_\gamma \ket{a_1} \bra{a_2}\  ]
\label{tr}
\ee
where the symbol $\tilde Tr$ means the trace over all matter 
+ gravity configurations
such that the WDW constraint is satisfied by integration over positive
 $N$.

To get some flavor of what
is the physical meaning of the ``Feynman'' boundary condition
in the present case,
it is appropriate to consider the
 the WKB approximation. 
In this approximation, the kernel is equal to  
\be
K (a_2, a_1 ; n_M, n_\gamma ) = { exp \left( i \int_{a_1}^{a_2}
\pi(a, n_M, n_\gamma)  da   \right) \over 2 \sqrt{ \pi(a_2, n_M, n_\gamma)
\pi (a_1, n_M, n_\gamma) }} = \Psi_{WKB} ( a_2; n_M, n_\gamma )
\Psi_{WKB}^* ( a_1; n_M, n_\gamma )
\label{M16}
\ee
where the momentum $\pi (a, n_M, n_\gamma)$ is
given in eq. (\ref{M8}), and where $\Psi_{WKB} ( a; n_M, n_\gamma )$
is the WKB solution of  eq. (\ref{M14}) with unit wronskian.
Therefore in that approximation, one has
monotonic expansion when $a_2 > a_1$ if the classical
momentum $\pi_a$ does not vanish. 
Then, ``negative energy'' solutions, with $\pi > 0$, are not 
generated since there is no back-scattering in the WKB approximation.
Upon considering matter interactions in \cite{wdwpt}, 
negative energy solutions might be produced. We refer to that 
paper for a discussion of this point.


In Section 5, we shall evaluate the corrections to the WKB
approximation. For the moment, we just recall
that its validity requires, 
see {\it e.g.} \cite{BD}, that 
\be
 \partial _a  \pi (a, n_M, n_\gamma) <\!\!<    \pi^2 (a, n_M, n_\gamma)
\label{M16'}
\ee
The reason for which we recall this criterion is the following: we 
want
to {\it compare} the implications of eq. 
(\ref{M16'}) and eq. (\ref{M1}).
To this end
we introduce the dimensionless quantity $N_s$,
given by $N_s(a) = 2G E_{m}(n_M, n_\gamma, a) / a $ which 
may be understood as the ``Schwarzschild radius'' of the matter content 
of the universe measured in the unit of $a$.
For a closed universe, one has $N_s = 1$  at its turning point. 
More generally, in (non-empty) cosmology, one is confronted with ``huge''
numbers, i.e. numbers whose logarithm are much bigger than one,
as well as large numbers, i.e. numbers whose logarithm are bigger but
close to one. What we want to stress is that
$N_S$  
belongs to the second class.
(We shall consider separately the de Sitter case.) Thus upon dealing 
with ``huge'' numbers, like $n_M$, one can put, in a first approximation, 
$N_S=1$ and trade therefore 
$a$ for the total {\it matter} content of the universe.

To see how this works, 
consider a matter dominated universe with $
N_s(a) = {\cal{O}}(1) $ and with $N_s(a) >\!\!> \Lambda a^2$.
Then, the condition that the Hubble radius be much bigger than the
Compton wave length,
eq. (\ref{M1}), leads to a constraint on $n_M$, the
number of massive quanta,
\be
 n_M >\!\!>  {\hbar  \over G M^2} = { \mu_{Planck}^2 \over M^2 }>\!\!>1
\label{M20}
\ee
Indeed one has $(\dot a /a)^2  \simeq G M n_M /a^3 \simeq (G  M n_M)^{-2}$
 when $N_s = {\cal{O}}(1)$. 
The last inequality in eq. (\ref{M20}) simply recalls
that $M$ must be much smaller than the Planck mass $\mu_{Planck}$
in order to legitimize the neglection of the gravitational dressing of
{\it individual} quanta.  This requirement has nothing to do with the 
semi-classical of the propagation of the cosmological radius $a$.
In this respect, it
 should be pointed out that the condition (\ref{M20}), is not
compatible with the limit $G \to 0$ which is used 
in refs. \cite{banks}\cite{kiefer2}\cite{ortiz} to define the background solution.
This is related to the fact that we shall find, in contradiction to these
works, that the background solution is driven by the mean matter energy. 

Using the same substitutions, eq. (\ref{M16'}), gives
\be
 n_M >\!\!>  { \mu_{Planck}\over M }
\label{M20'}
\ee
When eq. (\ref{M20}) is satisfied,
 eq. (\ref{M16'}) is fulfilled.
This simply means that when the Hubble radius $a/\dot a$
 is much bigger than the
Compton wave length, matter dominated universes behave semi-classically.
We emphasize this point: the validity of the WKB approximation
for $a$ puts constraints on the {\it sources} of gravity, i.e. on matter.
Furthermore, this hierarchy of constraints on $n_M $ determines
 the {\it relative} importance of the corrections on the
various approximations that we shall perform in the next section.
This will be made explicit in Section 5.

\section{Green functions from WDW solutions}

To construct Green functions from the 
solutions of the WDW equation,
we shall proceed in two steps.
First, we work in the background field
approximation, i.e. we
evaluate the Green functions of $\psi$ and $\phi$ from their time
dependence
in a classical universe whose radius follows $a= a(t)$.
Secondly, we define an extension of these Green functions
in the WDW framework and show under which
conditions the quantum matter-gravity evolution leads to 
the expressions previously obtained.

In a classical universe whose radius is given, once for all, by $a= a(t) $,
the Green functions of the $\psi$ field are defined
 in total analogy with eq. (\ref{three}), 
by
\be
G_{\psi} (t_2, t_1) = \expect{{\chi }_0}
{\psi (t_2) \psi (t_1) }{\chi_0}
\label{threepsi}
\ee
where $\ket{\chi_0}$ is the state of the $\psi$ field.
When the state of the field is described by a diagonal density
matrix, one obtains
\ba
G_{\psi}^{\rho_M} (t_2, t_1) &=&
Tr_\psi [  \hat \rho_M \  \psi(t_2) \psi(t_1)  ] 
=
 \Sigma_{n_M} \rho_{n_M} 
G_{\psi}^{n_M}(t_2-t_1)
\nonumber\\ 
&=&
 \Sigma_{n_M} \rho_{n_M} {1 \over 2 M}  \left(
{n_M} \ e^{i M (t_2-t_1)}  + ({n_M}+1) \ e^{-i M (t_2-t_1)}
 \right)
\label{sixpsi}
\ea
where the symbol $Tr_\psi$ has been introduced to be used in 
concordance with eq. (\ref{tr}). It means ``trace over the $\psi$ 
configurations''.
The last equality 
follows from the fact that the $\psi$ field is an harmonic
oscillator of frequency $M$ for all cosmological histories described
by $a(t)$, see eqs. (\ref{fur}, \ref{six}).
For the $\phi$ field instead, one gets
\be
G_{\phi}^{\rho_\gamma}  (t_2, t_1; k ) = \Sigma_{n_\gamma} \rho_{n_\gamma}
  {1 \over 2 k} \left(
{n_\gamma} \ e^{i k (\eta_2-\eta_1)}  + ({n_\gamma}+1) \ 
e^{-i k (\eta_2-\eta_1)}
 \right)
\label{sixphi}
\ee
where the lapse of conformal time is defined by
$\eta_2-\eta_1 = \int_{t_1}^{t_2} dt/a(t)$.

We now define a matrix element which generalizes eq. (\ref{sixpsi}) 
in the WDW framework.
In this case as well, we have to specify the matter content of the 
universe, i.e. the weights $\rho_{n_M}, \rho_{n_\gamma} $.
However we shall see that the matter content will now play a 
{\it double} role,
the former one appearing in eq. (\ref{sixpsi}) and the one of eq.
(\ref{M9}) which
determines the time lapse as a function of $a_2$ and $a_1$.

Using the notation presented in eqs. (\ref{tr}, \ref{sixpsi}),
we define ${\cal{G}}_{\psi}^{\rho} (a_2, a_1) $ to be
\be
{\cal{G}}_{\psi}^{\rho_\gamma} (a_2, a_1) =
 \tilde Tr \left[ \ \hat \rho_M \hat \rho_\gamma 
\left(
\ket{a_2} 
i\!\!\lr{\partial_{a_2}}
\psi
\bra{a_2} \right) \ 
\left(
\ket{a_1} 
i\!\! \lr{\partial_{a_1}}
\psi
\bra{a_1} \right) \ \right]
\label{G}
\ee
The operator $\psi$ is the Schroedinger operator $= (d_M + d_M^\dagger)/
\sqrt{2M}$. 
Then the operator $\ket{a_2} i \!\!\lr{\partial_{a_2}}
\psi \bra{a_2}$ should be read  ``$\psi$ at $a=a_2$''.
It replaces exactly the
Heisenberg operator $\psi(t_2)$ in eq. (\ref{sixpsi}) which is 
``$\psi$ at $t=t_2$''.
The occurrence of the operator $2 \pi_{a_2} =  i\!\! \lr{\partial_{a_2}}$
follows from the second order character of the propagation of $a$,
see the discussion after eq. (\ref{M15}).
We shall see that this operator guarantees that the weight 
attributed to ``to be at $a=a_2$''
does not depend on the momentum $\pi_a$ at $a_2$,
 hence ``to be at $a=a_2$''
 is independent of the matter content of the universe\footnote{
The origin of this operator $2 \pi_a$ can also be understood
as follows. Green functions are densities with respect to $t$ (or $a$)
since both are continuous variables. This means that 
Green functions enter
into physical rates of transitions integrated over $t$ (or $a$).
Therefore instead of considering ${G}_{\psi}^{\rho} (t_2, t_1)$
one might consider ${G}_{\psi}^{\rho} (t_2, t_1) dt_2 dt_1$ directly.
Then, the corresponding quantity in quantum cosmology
becomes ${\cal{G}}_{\psi}^{\rho_\gamma} (a_2, a_1) da_2 da_1 / 
\pi_{a_2}\pi_{a_1}$ and the extra factors of $\pi_a^{-1}$
cancel with the previous ones. Upon considering {\it interactions},
it is the hamiltonian framework which specifies the
weight accompanying Green functions in transition probabilities,
and indeed one finds that
unit wronskian is required\cite{wdwpt}. However extra factor of $1/
\sqrt{2|\pi_a|}$ will still be found at the end points
of the amplitudes. Their
amputation is mandatory in order to obtain amplitudes of 
{\it probability},
as in the reduction formula.}
i.e. independent of $n_M, n_\gamma$. This independence is 
required in order to interpret the matrices $\hat \rho$ of eq.
(\ref{G}) as providing the {\it probability} to
find the system characterized by $n_M, n_\gamma$, being at $a=a_2$.

In terms of the kernel at fixed matter content, see eq. (\ref{M15}),
one gets
\ba
{\cal{G}}_{\psi}^{\rho} (a_2, a_1) =
 \Sigma_{n_M, n_\gamma} \rho_{n_M}   \rho_{n_\gamma}
\left[ {n_M \over 2 M} 
K(a_2, a_1 ; n_M - 1 , n_\gamma ) 
 + 
{ n_M + 1 \over 2 M}  
K(a_2, a_1 ; n_M + 1 , n_\gamma ) 
\right] &&
\nonumber\\ 
i\!\! \lr{\partial_{a_2}} i\!\! \lr{\partial_{a_1}} K(a_2, a_1 ; n_M, n_\gamma )
\quad \quad  \quad \quad \quad \quad  \quad \quad  \quad \quad 
 \quad \quad  \quad \quad &&
\label{KK}
\ea
where the two terms come, as in eq. (\ref{fur}), from 
the products $d_M^\dagger d_M$ and the $d_M d_M^\dagger$ which  decrease
and { increase} respectively the number of massive quanta by {\it one} unit.

Since 
$n_M$ and $n_\gamma$ are strictly conserved, ${\cal{G}}^{\rho}$ 
decomposes as
\ba
{\cal{G}}_{\psi}^{\rho} (a_2, a_1) =
 \Sigma_{n_M, n_\gamma} \ \rho_{n_M}   \rho_{n_\gamma} \  
{\cal{G}}_{\psi}^{n_M, n_\gamma} (a_2, a_1) 
\label{sum}
\ea
where ${\cal{G}}_{\psi}^{n_M, n_\gamma} (a_2, a_1) $
is the Green function in a universe with exactly $n_M$ and $n_\gamma$ quanta.
This is our first result.

We now proceed to a {\it double} development.
First, we use the WKB expression for $K(a_2, a_1 ; n_M, n_\gamma )$,
eq. (\ref{M16}), and we obtain (compare with eq. (\ref{fur}))
\ba
{\cal{G}}_{\psi}^{n_M, n_\gamma} (a_2, a_1) &=&
{n_M \over 2 M}  
  \  A_{-}\ 
e^{-i \int_{a_1}^{a_2} [ \pi(a, n_M, n_\gamma) - \pi(a, n_M- 1, n_\gamma) ] da}
\nonumber\\
&&
+ { n_M + 1  \over 2 M}
A_{+}\ 
e^{-i \int_{a_1}^{a_2} [ \pi(a, n_M, n_\gamma) - \pi(a, n_M +1 , n_\gamma) ] da}
\label{WKKB}
\ea
where $\pi(a, n_M, n_\gamma) $ is the classical momentum, solution
of eq. (\ref{M8}), and where the normalization factors $A_{\pm}$ are given by
\be
A_{\pm} = {1 \over 4} {
\left[ \pi(a_1, n_M, n_\gamma) + \pi(a_1, n_M \pm 1, n_\gamma) 
\right] 
\left[ \pi(a_2, n_M, n_\gamma) + \pi(a_2, n_M \pm 1, n_\gamma) 
\right]
\over 
\left[
 \pi(a_1, n_M, n_\gamma) \ \pi(a_2, n_M, n_\gamma) \  
\pi(a_1, n_M \pm 1, n_\gamma)
\  \pi(a_2, n_M \pm 1, n_\gamma) \right]^{1/2}
}
\label{pis}
\ee

Secondly, we develop the phase and the norm 
of ${\cal{G}}_{\psi}$ to first order in the {\it change} 
of the matter energy ($=M$),
that is to the zeroth order in $1/ n_M$. 
Using  Hamilton-Jacobi relation, the phase can be written as
\ba
\int_{a_1}^{a_2} [ \pi(a, n_M, n_\gamma) - \pi(a, n_M- 1, n_\gamma) ] da &=&
\partial_{n_M} \int_{a_1}^{a_2} \pi(a, n_M, n_\gamma) da + {\cal{O}}(1/n_M)
\nonumber\\
&=& -
M \ \Delta t ( a_2, a_1 ; n_M, n_\gamma) + {\cal{O}}(1/n_M)
\label{dt}
\ea
by definition of 
$\Delta t ( a_2, a_1 ; n_M, n_\gamma)$, c.f. eq. (\ref{M9}).
To this order in $1/n_M$, one finds that the norm 
$A_{\pm} =1 $, for all $n_M, n_\gamma$.
This is the necessary condition mentioned above. The 
gravitational propagation must be normalized in such a way that it
gives a unit weight to the matter matrix elements, at least to 
this order in $1/n_M$.
  
Collecting the results, one gets 
\ba
{\cal{G}}_{\psi}^{n_M, n_\gamma} (a_2, a_1) &=& {1 \over 2 M} 
\left ( n_M \ e^{i M \Delta t ( a_2, a_1 ; \ n_M, n_\gamma) }
+ ( n_M + 1 ) 
e^{-i M \Delta t ( a_2, a_1 ; n_M, n_\gamma) }
\right)
\nonumber\\
&=& G_{\psi}^{n_M}(\Delta t ( a_2, a_1 ; \ n_M, n_\gamma))
\label{GG}
\ea
see eq. (\ref{sixpsi}).

Therefore, we have proven that to first order in the change
in the matter energy and to the WKB approximation
of the kernels for gravity, 
$ {\cal{G}}_{\psi}^{n_M, n_\gamma} (a_2, a_1)$, defined in eq.
(\ref{G}), 
is {\it equal} to the Green function
 $G_{\psi}^{n_M}(\Delta t ( a_2, a_1 ; \ n_M, n_\gamma))$ evaluated
in {\it the} universe characterized by $n_M$ and $ n_\gamma$. 
This is our second result. 
From eq. (\ref{WKKB}), we see that the sole role
of the product of the two normalized kernels 
$K( a_2, a_1 ; \ n_M, n_\gamma)$ 
is to provide the phase of the Green function
$G_{\psi}^{n_M}(\Delta t ( a_2, a_1 ; \ n_M, n_\gamma))$.

The double dependence in $n_M$, that we mentioned after eq. (\ref{sixphi}),
is now manifest. As in eqs. (\ref{fur}, \ref{sixpsi}), there is the
usual ``harmonic'' dependence of the norm 
in the number of quanta under examination, 
{\it i.e.} those of the $\psi$ field,
but there is also the {\it parametric} dependence of the lapse
of proper time $\Delta t ( a_2, a_1 ; \ n_M, n_\gamma)$ in
the total number of quanta of all species that are present in the 
universe. For exactly the same reason,
i.e. the dynamical character of gravity,
a double dependence of the temperature in the 
occupation numbers is also found upon
considering statistical mechanics in the presence of gravity,
see the discussion after eq. (25) in \cite{PKO}. 

We emphasize that the quantity appearing in the
Green functions, i.e. $M \Delta t = \partial_n \int da \pi_a$
is gauge independent. Indeed, it gives the number of nodes of
the Compton frequency $M^{-1}$ from $ a_2$ to $ a_1$ in a universe
filled with $n_M$ and $ n_\gamma$ quanta. The 
reason which guarantees this gauge invariance is that 
(i)
the ``entries'' of 
the kernels $K( a_2, a_1 ; \ n_M, n_\gamma)$
 are the gauge independent quantities
$a$, $n_M$ and $ n_\gamma$, (ii)
the Green functions are determined in terms
of variations of $K$ with respect to these
quantities.

We also emphasize that the decomposition in eq. (\ref{sum})
requires only the existence of conserved quantities, 
in our case $n_M$ and $n_\gamma$\footnote{
We shall relax the constraint on the Hubble radius, 
eq. (\ref{M1}), in \cite{wdwpc}
 and consider pair creation of massive quanta.
Then exactly conserved quantities will be replaced by adiabatically
conserved quantities.}.  No other approximation is needed. 
Eq. (\ref{sum}) is essential in guaranteeing that the background
geometry, from which time lapses
are defined, is the one characterized by these quantum numbers.
In this we confirm and generalize  
\cite{HalH}\cite{vil}\cite{BV} and disagree with
 \cite{banks}\cite{kiefer1}\cite{kiefer2}\cite{ortiz}.
Indeed there is no reason to develop 
eq. (\ref{dt}) around $n_M=1,  n_\gamma = 0$.
It is this procedure which is (implicitly) adopted in those works
by taking the limit $G \to 0$ in eq. (\ref{M14}).

In this respect, it is very instructive to compare the parallel
development which leads to a canonical 
distribution for a little system contained in a much bigger isolated ensemble. 
First, one realizes that the action  of the cosmological expansion, 
eq. (\ref{M12'}),
acts like the entropy of a reservoir in delivering the time lapse. Indeed, 
a heat reservoir determines the (inverse) temperature 
through a first order energy change of its entropy,
compare eq. (\ref{dt}) and eq. (\ref{betab}) in Appendix B. 
Secondly, upon considering second order energy changes,
the analogy is maintained and reinforced since 
the corrections to the background field approximation 
are similar to to the finite size effects of a big isolated system, see 
Section 5. Thirdly, the additive character of the energy
implies that two different massive fields evolve with the same proper time.
 Indeed,
 to first order in the energy changes, the derivative of eq. (\ref{dt}) 
would determine the same time lapse for both fields. This is exactly like
the ``zero-th law'' of thermodynamics 
(i.e. the equality of the temperatures at equilibrium) which
follows in a microcanonical ensemble from the additivity of the
energy and the equipartition ansatz.   

\vskip .3 truecm
\noindent
Before proceeding to the evaluation of the corrections
and exploring the correlated nature of both approximations used,
three important remarks should be added.

1. Had we studied the corresponding Green function for the
massless conformally coupled field $\phi$, we would have obtained 
\ba
{\cal{G}}_{\phi}^{n_M, n_\gamma} (a_2, a_1 ; k) &=& {1 \over 2 k} 
\left ( n_\gamma \ e^{i k \Delta \eta ( a_2, a_1 ; \ n_M, n_\gamma) }
+ ( n_\gamma + 1 ) \
e^{-i k \Delta \eta ( a_2, a_1 ; n_M, n_\gamma) }
\right)
\label{GGeta}
\ea
instead of eq. (\ref{GG}). This is because the gravitational phase in the WKB 
approximation is governed by the change of $\int \pi da$ with respect of 
$n_\gamma$:
\be
\partial_{n_\gamma} \int_{a_1}^{a_2} \pi(a, n_M, n_\gamma) da = -
k \ \Delta \eta ( a_2, a_1 ; n_M, n_\gamma)
\label{deta}
\ee
by definition of $\Delta \eta ( a_2, a_1 ; n_M, n_\gamma)$, see eq. (\ref{M10}).
Thus, the response of gravity, that is the first order change 
of $\int \!da \ \pi(a, n_M, n_\gamma)$
induced by the operators $\psi$ or $\phi$, furnishes (by definition) the correct
time dependence through Hamilton-Jacobi equations.

2. Eq. (\ref{sum}) 
tells us that the Green function is given by the weighted sum of
Green functions at fixed $n_M, n_\gamma$.
Therefore 
one has
\ba
{\cal{G}}_{\psi}^{\rho} (a_2, a_1) &=&  
\Sigma_{n_M, n_\gamma} \ \rho_{n_M}   \rho_{n_\gamma} \ 
G_{\psi}^{n_M}(\Delta t ( a_2, a_1 ;  n_M, n_\gamma))
\nonumber\\
&=& 
\Sigma_{n_M}\  \rho_{n_M}  
\ G_{\psi}^{n_M}(\Delta t ( a_2, a_1 ;  \bar n_M, \bar n_\gamma))
=
G^{\rho_M}_{\psi} ( \bar \Delta t ( a_2, a_1 ) )
\label{GG3}
\ea
where $\bar n_M, \bar n_\gamma$ are the mean number 
of massive and massless quanta
and where the mean lapse $\bar \Delta t ( a_2, a_1 ) $ is equal to 
$\Delta t ( a_2, a_1 ; \ \bar n_M,  \bar n_\gamma)$.
In the second equality, we have replaced the {\it parametric} dependence
of ${\cal{G}}$ through $\Delta t( a_2, a_1 ; n_M, n_\gamma)$
in $n_M$ and $n_\gamma$ by a single  ``background'' contribution 
controlled by the mean occupation numbers.
This is physically relevant for well peaked distributions,
i.e. for $\bk{(\Delta n)^2}/\bk{ n^2}  <\!\!< 1$. In that case,
the Green functions ${\cal{G}}$ are correctly given in terms of the properties
of the {\it mean} universe only.
This mean universe is the solution of Einstein equations 
driven by the mean energy density\cite{brout}\cite{vil}\cite{BV}
(the so-called semi-classical equations). In our case, it is determined by,
see eq. (\ref{M6}),
\ba
 H_G + \bk{H_{m}}&=& 0
\nonumber\\
 - { \dot a^2 \over a^2 }+ {\kappa \over a^2 } + \Lambda 
 + { 2G \over a^3 } \left( M \bar n_M + { k \over a }
\bar n_\gamma \right)&=& 0
\label{meaneq}
\ea
where $\dot a$ means $\partial_{\bar t} a $, i.e. derivative with respect to the 
mean proper time.

Instead, for widely spread distributions, it is no longer 
meaningful to define a mean universe 
governing mean lapses because remote matter field configurations
will classically evolve differently and quantum mechanically
will never interfere. However this is free of 
consequences since 
{\it each} contribution at fixed $n_M, n_\gamma$ 
delivers a well defined time dependence to Green functions.
We shall return to these ``decoherence'' aspects upon 
considering interactions, in \cite{wdwpt}.


3. We stress that eq. (\ref{WKKB}) delivers the Green function in
the Schroedinger picture, see eq. (\ref{fur}).
We now understand that the double dependence on 
$n_M$ and the parametric dependence in $n_\gamma$
lead to non-linearities which prevent to 
define an Heisenberg operator $\psi(a)$ from $H_M$ only.
On the contrary, the superposition principle still survives for
the Schroedinger evolution of the full wave function, here
given by $K(a_2, a_1; n_M, n_\gamma)$.

%

\section{The gravitational corrections to Green functions}

There are three types of corrections associated with the
three approximations used.
First there are the nonlinear terms in the change of the matter energy
induced by the operators $\psi$, see the passage from
eq. (\ref{WKKB}) to eq. (\ref{GG}). Then one has the corrections
to the WKB approximation used when replacing the kernels
in eq. (\ref{KK}) by their WKB expression given in eq. (\ref{WKKB}).
Finally one has the corrections associated with the spread in
$ n_M$ and $ n_\gamma$ upon computing the
Green function in the mean universe, see eq. (\ref{GG3}).
The first two are of dynamical character while the third one
depends of the choice of the state of matter, i.e. the spread 
in $\hat \rho_M$ and $\hat \rho_\gamma$.
Our aim is to determine what are the dominant and 
hence relevant corrections.

These three corrections make explicit the non-linearities
induced by the dynamical character of gravity. In particular,
the approximate nature of the background is displayed and its 
sensitivity on the quantum transition under examination is pointed out.

We start by evaluating the non linear corrections in the energy change 
which arise in the
phase and the norm of eq. (\ref{WKKB}).
To second order in the energy change,
the phase of the first term gives
\ba
\int_{a_1}^{a_2} \left[ \pi(a, n_M, n_\gamma) - \pi(a, n_M- 1, n_\gamma) 
\right] da =\quad \quad\quad \quad\quad \quad \quad \quad \quad &&
\nonumber\\
\int_{a_1}^{a_2} \left[ { a M \over G \pi(a, n_M, n_\gamma) }
\left( 1 - { a M \over G \pi^2(a, n_M, n_\gamma) } \right)\right] da 
\simeq \quad \quad \quad \quad \quad \quad &&
\nonumber\\
 - M \Delta t( a_2, a_1 ; n_M, n_\gamma) 
\left[ 1 - { a_1 M \over G \pi^2(a_1, n_M, n_\gamma) } 
\left( 1 +  \Delta t
({ \dot a \over a } -
 {2 \dot \pi
\over \pi
}) \right)
\right]  &&
\label{C1}
\ea
Notice that we have developed the expression around $n_M$ 
and not around the ``mean'' value $n_M - 1/2$. We have developed
around $n_M$ 
since we want to use the same background for both terms 
of eq. (\ref{WKKB}).

In the third line we have replaced the integral of the 
correction by the value of the integrand at $a_1$ times the interval 
plus the linear dependence in $ \Delta t$.
This is valid if $a / \pi^2$ is slowly varying.
To estimate these terms, one should specify the dominant
matter content of the universe.
When the universe is matter dominated, the first correction term is 
\be
{ a_1 M \over G \pi^2(a_1, n_M, n_\gamma)} 
\simeq {M  \over 2 M n_M + 2 k n_\gamma /a }  <  {1 \over 2 n_M }
\label{C2}
\ee
It is independent of $G$, $\hbar$ and $a$. It depends only on the
ratio of the matter change ($=M$) and the total energy in the universe.
This is what guarantees the validity of the background field approximation:
that the {\it sources } of gravity in the universe be heavy\footnote{As pointed out by 
R. Brout upon discussing
these results, the emergence of an inertial time
 delivered by the matter content of the universe can be viewed as 
rephrasing (and legitimizing) Mach's principle in this quantum context.}
when compared to the transition upon examination. This is our third result.

The origin and the meaning of this term are clear. It arises
from our choice to develop
eq. (\ref{C1}) around $n_M$ 
and not around the ``mean'' value $n_M - 1/2$. Indeed,
had one developed around the ``mean'' value, it would not be present.
Its meaning is therefore that 
each term of eq. (\ref{WKKB}) defines its own background. This
dependence of the background in the quantum event under examination
will be further discussed in \cite{wdwpt}.

As already pointed out in Section 4, there is a close correspondence
 between the
present development of the gravitational action and the development
of the entropy of a big isolated system around equilibrium. Compare
eq. (\ref{C1}) and eq. (\ref{n56}) in Appendix B. We reemphasize that both
first order and second order energy changes are in strict correspondence.
Indeed, first order changes determine time and inverse temperature
and second order changes determine the corrections to these
background-equilibrium concepts.

We emphasize that the Planck length does not appear in this correction 
and we point out that this correction has the same origin as the
recoil effect of a relativistic particle affected by a transition\cite{rec}.
In that case the correction term\footnote{In this respect, it is instructive to
compare the perturbative treatments  of \cite{kiefer2} 
and \cite{rec} which are both
 applied to a relativistic particle. Each treatment is
closely related to the corresponding one in quantum cosmology.
In \cite{kiefer2}, the development is made
around zero momentum (this corresponds to isolate empty Einstein solutions)
whereas in \cite{rec}  the development is made
around the {\it change} in the momentum induced by a given transition. 
This latter clearly corresponds to what was done in eq. (\ref{C1}).} 
is
$\om / 2 M$ where $\om$ is the energy of the emitted photon and $M$ the rest 
mass of the particle, more on this can be found in \cite{wdwpt}.

When the universe is radiation dominated, this correction term
is $ a M / n_\gamma k$
which is smaller than $1/ n_M$ and independent of $G$ as well.
When the universe is de Sitter dominated,
the correction is $ G M / \Lambda a^3 (<< 1/ n_M$ by hypothesis).

The second correction term of eq. (\ref{C1}) is quadratic in $\Delta t$. It
scales like the first correction term multiplied by $\Delta t (\dot a / a)$. 
Therefore as long as $\Delta t$ is of the order of a few Compton wave lengths this
second term is much smaller than the first one by virtue of eq. (\ref{M1}).
By a similar analysis, one easily shows that the norm factors
$A_\pm$, eq. (\ref{pis}) are given by $( 1 + 1/n^2_M {\cal{O}} (1))$.

Therefore, for macroscopic universe characterized by $n_M >> 1$,
the non linear terms engendered by the 
change in energy carried by the operators $\psi$
can be correctly and safely discarded.
Furthermore, the dominant correction comes from the desire of
working with a {\it single}
 background characterized by the initial number of quanta
$n_M$ hence valid for both terms of the Green function.

We now compute the corrections to the WKB approximation. We recall
that we 
used this approximation in order 
to evaluate the kernels of gravity $K( a_2, a_1 ; n_M, n_\gamma)$,
see eq. (\ref{M16}).
Very important is the fact that ${\cal{G}}$ is given in term
of products of two kernels, see eq. (\ref{KK}). Indeed
 only a small part of the correction to the $K$'s will therefore
affect the Green functions.

As explained in \cite{BD}, the first order correction
to the WKB approximation is obtained by replacing $\pi(a, n_M, n_\gamma)
= \pi_{cl}$
by
\be
\tilde \pi(a, n_M, n_\gamma) = \pi_{cl} 
\left( 1 +  {\hbar^2 \over 4 }
\left[ { 3 \partial_a \pi_{cl}^2 \over 2
\pi_{cl}^4 } - { \partial^2_a \pi_{cl} 
\over 
\pi_{cl}^3 } \right]
\right)
\label{C3}
\ee
As before, one has to specify the matter content of the universe to 
evaluate the modification of the phase and the norm of ${\cal{G}}$
induced by this correction. 
When the universe is matter dominated, the modification of
phase of the Green function induced by the second term of eq. (\ref{C3})
is, see eq. (\ref{C1}),
\ba
 - M \Delta t( a_2, a_1 ; n_M, n_\gamma) 
\left( { \mu^4_{Planck} \over M^4 } { 1 \over n_M^4 } {\cal{O}} (1)
 \right) <\!\!<
- M \Delta t( a_2, a_1 ; n_M, n_\gamma) 
\left( { 1 \over n_M^2} \right)
\label{C4}
\ea
where we have used eq.  (\ref{M20'}).
This means that when eq. (\ref{M1}) is satisfied, the universe 
behaves so semiclassically that the part
of the correction to the WKB approximation which modifies the 
Green function is {\it irrelevant} since it
is smaller
than the 
non-linear corrections
of eq. (\ref{C2}) by an extra factor of $1/ n_M$.

Similarly, the change of the phase of the Green function 
associated with a different choice of the ``normal ordering'' of the operator 
$\partial^2 _a  $ in eq. (\ref{M14}) is equally irrelevant.
Indeed another choice in defining $\partial^2 _a  $ would lead to 
an additional quantum potential term which would induce a 
change in  eq. (\ref{C1}) through a change in the momentum $\pi_a$
of the same order of the WKB corrections given in eq. (\ref{C3}).

In both cases, the relevant part of the 
corrections
is given by 
the {\it change} of the correction obtained by comparing two neighbouring
solutions characterized by $n_M$ and $n_M - 1$. Indeed 
both the phase and the
norm of ${\cal{G}}$ are given in terms of 
the difference $\pi(a, n_M, n_\gamma) -
 \pi(a, n_M- 1, n_\gamma)$. This is why small corrections
are obtained and why we emphasize the necessity of the 
{\it double} development. As in eq. (\ref{C2}), it 
is the weight of the {\it sources} of gravity which guarantees 
the negligible
character of these corrections.

\vskip .3 truecm

The third kind of corrections are associated with the spread
of $n_M, n_\gamma$ specified by the choice of the 
matrices $\rho_M, \rho_\gamma$.
To estimate these corrections and to compare them 
with the first kind just described, we
write the mean quadratic spread as ${(\bar\Delta n_M)^2} =  n_M \sigma_M$
and similarly for $\bar\Delta n_\gamma$.
Then by repeating the algebra of eq. (\ref{C1}),
one verifies
that the change in the phase is controlled by the factor
$1 + \sigma^2_M /\bar n^2_M {\cal{O}} (1)$. This result
was obtained in \cite{BV}.
Similarly,
the prefactor of ${\cal{G}}$ is multiplied by 
$1 + i M \Delta t \sigma_M / \bar n_M $.
Thus for sufficiently small time intervals,
the corrections induced by the spread are negligible for $n_M >> 1$
and $\sigma_M = {\cal{O}} (1)$.
Note however that the ``$1$'' in the factor $n_M + 1$ which
appears in the mean Green function defined in eq. (\ref{GG3})
 is meaningful only if $M \Delta t \sigma_M << 1$.

 Had we worked with a universe described by a pure state
instead of a diagonal density matrix, we would have obtained the
 additional interfering terms of eq. (\ref{five}). 
This interfering term
 is more sensitive to gravitational corrections. 
Indeed, for $t_2 + t_1 - 2t_0$ large compare to 
$\bar n_M / \sigma_M$, the
phase shifts due to the non-linearities introduce additional
phases which will lead to the vanishing of that term upon summing over 
$n_M$ and $n_\gamma$.
Therefore, the backreaction of gravity, i.e. the dependence of the 
time intervals in the particle content of the universe (which
is represented by these non linear terms),
leads inevitably to decoherence for sufficiently long time intervals.
All of this is valid for well peaked distribution of energy 
around their mean values. For wider spreads, the decoherence 
effects are even stronger. 
 We shall return to this point 
after have considered the interactions in \cite{wdwpt}.
Let us just mention that the interactions
among matter and radiation will inevitably engender
spreads in $n_\gamma$ proportional to $n_M$. Therefore
density matrices which are initially too peaked will
spread dynamically. Spreads in energy are intrinsic to cosmology.

Finally we stress that up to now the insertion of the
field operators at some radius has been performed from the outset
as if one had an {\it additional} quantum system coupled to the field
at that radius. However such an additional system 
carries some mass-energy which must be taken
into account in the total matter hamiltonian. It will therefore
affect gravity. 
To
show how to treat the interactions of the {\it internal}
constituents of the universe is the purpose of \cite{wdwpt}.

\vskip 1. truecm
{\bf Acknowlegdments    }

\noindent
I would like to thank R. Brout for three years of intense discussions on
the aspects developed in this paper, as well as for useful
comments concerning a first draft of this paper.
I am grateful to R. Balbinot, Cl. Bouchiat, J. Iliopoulos,
S. Massar, Ph. Spindel and G. Venturi for useful remarks.
I wish also to thank the group of the CNRS center of Marseille-Luminy
for an invitation during which various aspects of this work were 
discussed.

\section{Appendix A. 
\newline
Green functions with non-vanishing momenta}

In this Appendix we introduce an extension of the formalism
in which the momenta of matter no longer vanish.
This extention allows to compare the manner by which the notion of
spatial displacement arises from homogeneous solutions
to the manner we used to recover time lapses from stationary solutions.
Furthermore, when dealing with homogeneous universes,
this extention makes explicit the appearance of 
inertial coordinates as the ``angles''
conjugated to the conserved momenta, i.e. the ``actions''.
Finally, the dynamical character of gravity, i.e. the backreaction,
breaks the symmetry between the local representation and
the momentum representation which exists when quantum field theory is
considered in the absence of gravity. Indeed, gravity responds
to the  conserved momenta. This implies that the local
representation is an approximate description which
can only be defined a posteriori.

Upon considering non-homogeneous gravitational and matter fields
one has to deal with a set a four local constraints ${\cal{H}}^\mu(x)=0 $
in the place of the single scalar constraint, eq. (\ref{M6}). These local
constraints are enforced by four Lagrange field-multipliers, $N^\mu(x)$,
the local lapse and the local shifts.
Up to now, these constraints have only been taken into account 
partially. For instance, in \cite{HalH}, the action of gravity and matter
have been developed to quadratic order only. This leads to a 
hierarchy of constraints. Indeed, to that order in the
fluctuations around an homogeneous background, the non-homogeneous
part of the constraints ${\cal{H}}^\mu(x)$ is {\it linear} in the fluctuations
of matter and gravitational configurations. Instead, the homogeneous parts of the
temporal constraint, $H$, i.e. the zero momentum 
component of ${\cal{H}}^0 (x)$, 
is quadratic in these fluctuations. The reason is very simple,
the non-homogeneous
parts of the Lagrange multipliers, $N^\mu(x)$, have no ``background''
contribution; only the zero momentum 
component of $N^0 (x)$, i.e. $N$, has such.

This has an important consequence for us. Upon dealing
 around an homogeneous background with
matter fluctuations which leads to {\it quadratic} contributions
in ${\cal{H}}^\mu(x)$, one has to work without gravitational 
fluctuations if one develops the gravitational fluctuations
in the action to quadratic
order only.
Accordingly, one should ignore completely the inhomogeneous
part of the constraints. This quadratic approximation
means that the matter quanta do not interact directly among themselves.
However
they do contribute to the homogeneous background energy.

We shall work in this framework.
The action of our massive field is now
\be
S_M = {1 \over 2 } \int d^3p \ dt \left( \dot \psi_p ^2 - 
(M^2 + {p^2 \over a(t)^2 } ) \psi_p ^2 \right)
\label{M3A}
\ee
where the dimensionless quantity $p$ is the conserved
number of nodes of the field fluctuation.
For closed three surfaces, $p$ is an integer number.
At this level, this action can be envisaged as describing a collection 
of homogeneous excitations characterized by the $a$ dependent energy
\be
\Om (p, a) = \sqrt{M^2 + p^2 /a^2}
\label{disp}
\ee
Similarly the action of the massless conformally coupled field is
\be
S_\gamma =  {1 \over 2 } \int d^3k\ d\eta  \left(
(\partial_\eta \phi_k)^2 - k^2  \phi_k^2  \right) =
{1 \over 2 } \int d^3k { dt \over a } \left(   a^2
 \dot \phi_k ^2 - k^2  \phi_k^2  \right)
\label{M4A}
\ee
 
We can then proceed as in Section 3.
 The energy of gravity must annihilate the $a$-dependent
total energy of the matter:
\be
H_G + H_{m}= \left({
-G^2 \pi_a + \kappa a^2 + \Lambda a^4
\over 2Ga }  \right) + 
\left( \int d^3 p \ n(M, p) \Om( p, a)  + \int d^3 k \ n(\gamma, k)
{ k \over a } \right)
=0
\label{M6A}
\ee
Notice that the subtraction of the zero-point energy
is now mandatory since we have an infinity of fluctuating modes.
We emphasize that, being conserved quantities, $p$ and $k$
are gauge invariant quantities. Therefore, upon making variations
of the gravitational kernel with respect to them, one shall define
gauge invariant quantities as well. In particular, their
conjugate variables are ``angles'', i.e. they do not appear
in the Hamilton-Jacobi equation. This is how inertial
coordinates are introduced and legitimized
in the hamiltonian framework. 
Furthermore, we treat $ n(M, p)$ as 
constants of motion as well,
 i.e. we work in the adiabatic approximation. This is valid 
if eq. (\ref{M1}) is satisfied.
Therefore, the total wave function 
 can be expressed, as in eq. (\ref{M13}), as a sum of waves at fixed particle
number $n(M, p) , n(\gamma, k)$. The same
is true for the kernel $K(a_2 , a_1 ; n(M, p) , n(\gamma, k))$ 
defined as in eq. (\ref{M15}).
Then
we can define the ``WDW'' Green function at fixed $p$ 
in a universe characterized by the set
of occupation numbers $n(M, p') , n(\gamma, k)$ by
\ba
&&{\cal{G}}_{\psi}^{n(M, p') , n(\gamma, k)} (p, a_2, a_1) 
= \tilde Tr \left[ \hat \rho_{n(M, p') } \hat \rho_{n(\gamma, k)} 
\left(
\ket{a_2}
i\! \lr{\partial_{a_2}}
\psi_p
\bra{a_2} \right) \
\left(
\ket{a_1}
i\! \lr{\partial_{a_1}}
\psi_p 
\bra{a_1} \right) \ \right] 
\nonumber\\ && \quad\quad\quad = \left[
 { n(M, p) \over 2 \sqrt{ \Om (p, a_1)\Om (p, a_2) }}
K(a_2, a_1 ; n(M, p) - 1 ,  n(M, p' \neq p ), n(\gamma, k) ) \right.
\nonumber\\  
&& \quad \quad \quad \quad \left.  +
{ n(M, p)  + 1 \over 2 \sqrt{ \Om (p, a_1)\Om (p, a_2) } }
K(a_2, a_1 ;  n(M, p)  + 1 , n(M, p' \neq p ), n(\gamma, k) ) \right] 
\nonumber\\ &&\quad \quad \quad \quad \quad 
i\! \lr{\partial_{a_2}} i\! \lr{\partial_{a_1}} K(a_2, a_1 ; n(M, p') , n(\gamma, k) )
\label{KKA}
\ea
where the derivatives act on the kernels only.

Upon using the WKB expressions for the kernels $K$ and to first order
in the change in energy, one obtains 
\ba
{\cal{G}}_{\psi}^{ n(M, p') , n(\gamma, k) }(p, a_2, a_1)
&=& 
{n(M, p)\ e^{ i \int_{t_1}^{t_2}
dt' \Om (p , a (t'))} \over 2 \sqrt{ \Om (p, a_1)\Om (p, a_2) }}
+ {( n(M, p)  + 1 ) \ e^{- i \int_{t_1}^{t_2}
dt' \Om (p , a (t'))} 
\over 2 \sqrt{ \Om (p, a_1)\Om (p, a_2) } }
\nonumber\\
&=&
G_{\psi}^{n(M, p)} [p, \Delta t( a_2, a_1 ; n(M, p') , n(\gamma, k) )]
\label{GFp}
\ea
where $G_{\psi}^{n(M, p')} [p, \Delta t( a_2, a_1 ; n(M, p') , n(\gamma, k) )]$
is the Green function evaluated in a universe whose particle
content is determined by the $n(M, p') , n(\gamma, k)$.

We are now in position to determine to what extend 
one recovers the usual local representation.
One first introduces $x$, the conjugated to the wave number $p$, by
 defining the (local) field operator $\psi(x) = \int d^3 p \ e^{ipx} \psi_p$.
Then the Green function of these operators is
\ba
{\cal{G}}_{\psi}^{n(M, p') , n(\gamma, k)} (x_2, x_1 , a_2, a_1) 
&=& \tilde Tr \left[ \hat \rho_{n(M,p')} \hat \rho_{n(\gamma, k)} 
\ket{a_2}
i\! \lr{\partial_{a_2}}
\psi(x_2)
\bra{a_2} \
\ket{a_1} i\! \lr{\partial_{a_1}}
\psi(x_1)
\bra{a_1} \right] 
\nonumber\\
&=& \int d^3 p \ e^{ip( x_2 - x_1)}\ 
{\cal{G}}_{\psi}^{n(M, p') , n(\gamma, k)} (p, a_2, a_1) 
\label{enx}
\ea
It is a function of $x_2 - x_1$ only 
since $p$ is a conserved quantum number. 
However it is not a Lorenz invariant function since the 
cosmological radius is a time dependent function.

The validity of the local representation depends
on the space-time separation of the two field
insertions at $a_1, x_1 $ and $a_2, x_2 $. Indeed, the momenta 
which furnish the dominant (saddle point) contribution
to ${\cal{G}}_{\psi}^{n(M, p') , n(\gamma, k)} (x_2, x_1 , a_2, a_1) $
are centered around the solution of
\ba
\Delta x &=& \partial_p \int_{a_1}^{a_2} da 
\left[ \pi( a, n(M, p') , n(\gamma, k)) - \pi( a, n(M, p)-1 , n(M, p' \neq p), 
n(\gamma, k)) \right]
\nonumber\\
& = & \int_{t_1}^{t_2} dt \
\partial_p \Om(p, a) \left[ 1 + {\cal{O}}( p / n_M M)  \right]
\label{psad}
\ea
Therefore the correction term is negligible only
if the two points are not too close to the light cone defined
by $\Delta x = \Delta \eta(a_2, a_1 ; n(M, p') , n(\gamma, k) )$
and reached in the limit $p \to \infty$.
To compute the value of a massive Green function in 
quantum gravity {\it on the light cone }
is meaningless since the ``recoil'' of gravity destroys the background.

 The lessons of this Appendix are the following

1- There is a regime in which it is 
legitimate to keep only the constraint
eq. (\ref{M6A}) and 
the dispersion relation eq. (\ref{disp}) of individual quanta.
This regime correspond to homogeneous background 
driven by incoherent matter quanta, i.e. the state of the matter is such
that the expectation value of the
momentum operator ${\cal{H}}^i_{matter}(x)$ vanishes. 

2- In that regime, gravity is only sensitive to the various occupation
numbers $n(M, p)$ weighted by $\Om(p, a)$. Therefore, all
matter quantities are evaluated in 
the  momentum representation, at fixed $p$. Only then, one may introduce,
by Fourrier transform,
local representations of these quantities since the reaction of gravity 
is a nonlinear function of $n(M, p)$ and $\Om(p, a)$.

3- 
The local representation is 
inevitably an approximation. Indeed, in order to localize with 
arbitrary precision one needs to include arbitrary momenta and
thus arbitrary energies. But the equality between the 
WDW Green function ${\cal{G}}_{\psi}^{n(M, p') , n(\gamma, k)} (p, a_2 , a_1)$
and the corresponding Green function evaluated in the background requires
that the exchanged energy $\Om(p, a) $
 be much smaller that the background energy\footnote{
Moreover upon considering the set of local constraint, this 
restriction on $\Om$ might be re-enforced. We hope to return on that point 
in a next publication}.
Thus the localizability of an event is limited in time and in space
by this restriction on the exchanged energy.
Therefore, exactly like the fact that the time lapse is an
approximate concept, it appears that spatial displacements $x_2 - x_1$
have a secondary approximative status. The distance is a useful
concept
which can be introduced for convenience,
a {\it maquillage} to use the word of Banks\cite{banks}.
 This latter aspect will become more clear upon
considering, in the next paper\cite{wdwpt},
the interactions of an heavy particle and the recoils induced 
by emissions of light quanta. Indeed, without interactions,
there is no physical justification to
introduce the local operator $\psi(x) = \int d^3 p \ e^{ipx} \psi_p$.

\section{Appendix B. \ 
The corrections in $v/V$ in a microcanonical ensemble}

In the Appendix we shall compute the induced (almost canonical)
partition function of a small system pertained in a bigger one. The total system 
 is microcanonically distributed.
For simplicity, we shall suppose that the both systems are homogeneous
and that the contact energy can be neglected.
The total system has a volume $V+v$ and the little system has a volume $v$.

The goal of this exercise is to determine how to obtain a well
defined expansion in $v/V$. This is not trivial and bears many resemblance
with the debated extraction of the background in quantum cosmology.

The total system has an energy $E$. Thus the number of 
states (to be more precise, the density of states)
can be expressed as 
\be
\Om_{total} (E) = \int_0^{E} d\e \ \Om_V (E - \e ) \ \om(\e)
\label{B11}
\ee
where $\Om_V (E - \e )$ is the number of states when the rest of the
big system with volume $V$ has an energy $E - \e $ and where $ \om(\e)$
is the number of states of the little system with energy $\e$.

Two alternative ways to evaluate this quantity will be considered and compared.
The first one consists in neglecting completely the 
energy of the little system (this seems legitimate since 
the ratio of the mean energies $\bar \e / (E- \bar \e )$ scales like $v/V$)
and in developing around the {\it vacuum} solution $\e = 0$. 
To first order in the change of $\Om_V (E - \e )$ in $\e$, one obtains
\ba
\Om_{total; 1} (E) &\simeq& \Om_V (E) \int_0^{E} d\e \ e^{- \beta_1 \e} \om(\e)
\nonumber\\
&=& \Om_V (E) \ z(\beta_1) 
\label{B12}
\ea
where the inverse temperature $\beta_1$ is defined by 
$\beta_1 = - \partial_\e \ln(\Om_V (E- \e))$ evaluated at $\e = 0$ and where
$z(\beta)$ designates the canonical partition function of the small system.

The second method consists in taking into account the the {\it mean}
energy of the little system $\e = \bar \e $ determined by the 
saddle point of eq. (\ref{B11}). This saddle expresses, as usual,
the equality of the temperatures:
\be
\partial_\e \ln \Om_V (E - \e ) + \partial_\e \ln\om(\e) = 0
\label{B13}
\ee
where 
\be
\beta_2=  \partial_E \ln\Om_V (E - \bar \e ) 
\label{betab}
\ee
The derivative is now evaluated 
at the saddle point energy $\e = \bar \e $. 
Then, to first order in the {\it change} in energy, i.e. $\e - \bar \e$,
 the partition function for the little system reads 
\ba
\Om_{total, 2} (E) &\simeq& \Om_V (E- \bar \e) 
\int_0^{E} d\e \ e^{- \beta_2 ( \e - \bar \e)}\  \om(\e)
\nonumber\\
&=& \Om_V (E - \bar \e) \ e^{ \beta_2 \bar \e} \ z(\beta_2) 
\label{B14}
\ea

To this {\it linear} order in the development of the density of the
big system $\Om_V$, the only difference between the two approaches
lies in the difference
in the temperatures only. Furthermore, one finds 
that $\beta_1 - \beta_2$ scales like $v/V$.
Therefore if one takes the thermodynamical limit at this moment, i.e.
$E \to \infty, V \to \infty$ with $E/V$ fixed, both methods give the same results.

However, this is no longer the case
if one wants to evaluate finite size effects, that is 
the fact that the big system has a finite heat capacity. Then, 
the difference between the two developments show up to second order
in the energy fluctuation. (This is inevitable
since $\beta_1 - \beta_2$ is of the order of $v/V$.)
Indeed, by developing $\Om_V(E-\e)$ to second order
in $\e$, one has 
\be
\Om_{total; 1} (E) = \Om_V (E) \int_0^{E} d\e \ e^{- \beta_1 \e} \
e^{- \beta_1^2 \e^2/ 2 C_V} \om(\e)
\label{n55}
\ee
in the first case and
\be
\Om_{total; 2} (E) = \Om_V (E- \bar \e) 
\int_0^{E} d\e \  
e^{- \beta_2 ( \e - \bar \e)} 
\ e^{- \beta_2^2 
(\e - \bar \e)^2/ 2 C_V} \om(\e)
\label{n56}
\ee
in the second case. We have neglected the difference in the heat capacity
of the big system
(defined by $\partial^2_\e \ln \Om(E-\e) = - \beta^2 / C_V$)
which is  induced by the different ``background'' energies $E$ and $E- \bar \e$,
since $C_V$ appears in the correction term only.

It is convenient to introduce the entropies $s = \ln \om(\bar \e),\ S = \ln \Om(E -\bar \e)$ 
as well as their relations with the mean energies: $s = \xi_1 \beta \bar \e = \xi_2 v \beta^3$
and  $S = \xi_1 \beta (E - \bar \e) = \xi_2 V \beta^3$. In those equations, the factors
$\xi_i$ are dimensionless and of the order of $1$.
It is also convenient to relate the specific heats to the entropies of both systems
as $C_V =  \xi_3 S$ and 
$ c_v =  \xi_3 s = \bk{ (\e - \bar \e )^2} \beta^2$. 
Then one finds that the mean correction in the first case scales like
\be
\bk{ \e^2 } {\beta^2 \over  C_V } = { s^2 \over S} { 1 \over \xi_1^{2}  \xi_3 } = 
{ v^2 \beta^3 \over V} { \xi_2 \over \xi_1^{2}    \xi_3 }
\label{rat1}
\ee
while in the second case it scales only like 
\be
 \bk{ (\e - \bar \e )^2}  {\beta^2   \over  C_V } = { c_v \over  C_V } 
= {s \over S} = { v \over V}
\label{rat2}
\ee
These results follows from the fact that, while the temperature is 
essentially fixed 
by the big system, it is the little one which determines, at a given temperature,
what are the mean quadratic fluctuations, i.e. the width around
the saddle energy $\e$. Notice that there is no need to differentiate the
temperatures to that order in $v/V$.

By inspection of eqs. (\ref{rat1}, \ref{rat2}), one sees
that only the second development is really a power series
in $v/V$. Instead, in the first case, the corrections contains the
 factor $v^2 \beta^3 / V$ which is ambiguous because it can be 
much bigger or much smaller than one when $v<< V$. 
Indeed, it suffices to consider a 
system such that $s << S$ but with 
$s^2 > S$.

The lesson is that when one wants to compute finite size effects,
i.e. first order correction in $v/V$, it is mandatory to take the mean energy
of the little system into account upon 
evaluating the temperature. the zeroth order ``classical'' solution which 
corresponds to the background from which the temperature is computed.
The parallelism between
these developments and the one appearing in eqs. (\ref{WKKB}, \ref{C1}) is 
manifest. 
*corresponds  zeroth order ``classical'' solution which 
%
In this respect we want to point out that the factor
$e^{\beta_2 \bar \e}$ of eq. (\ref{B14}) finds its perfect counterpart 
in ref. \cite{BV}
wherein the Schroedinger time ($\beta$) dependent equation differs
from the usual one by $e^{i t \bar \e}$. Its origin is identical
to our, i.e.  the background (equilibrium) is determined
through a saddle point which does involve the mean ``matter'' energy $\bar \e$.


\begin{thebibliography}{999}

\bibitem{isham} C. J. Isham, {\it Canonical Quantum Gravity and the Problem
of the time}, In ``Integrable Systems, Quantum Groups and Quantum Field Theories''
Kluwer Academic Publishers, London 1993. see also gr-qc/9210011
\bibitem{hartle} J. B. Hartle, in ``Gravitation and Cosmology'' edited by B.
Carter and J. Hartle, Plenum Press (1986)
\bibitem{ontime} R. Brout, {\it On time}, Proceedings to the First Sakharov 
Conference (1991)
\bibitem{rub}  V. Lapchinski and V. Rubakov, Acta Phys. Pol. B10 (1979) 1041
\bibitem{banks} T. Banks, Nucl. Phys. B249 (1985) 332
\bibitem{HalH} J. Halliwell and S. Hawking, Phys. Rev. D31 (1985) 1777
\bibitem{brout} R. Brout, Found. Phys. 17 (1987) 603
\bibitem{Hal} J. Halliwell, Phys. Rev. D36 (1987) 3626
\bibitem{vil} A. Vilenkin, Phys. Rev. D39 (1989) 1116
\bibitem{BV} R. Brout and G. Venturi, Phys. Rev. D39 (1989) 2436
\bibitem{kiefer1} C. Kiefer and T. Singh, Phys. Rev. D44 (1991) 1067
\bibitem{kiefer2} C. Kiefer, {\it The semiclassical approximation
to quantum gravity}, in ``Canonical Gravity-from Classical to Quantum''
ed. J. Ehlers and H. Friedrich (Springer, Berlin 1994), gr-qc/9312015
\bibitem{ortiz} G. Lifschytz, S. Mathur and M. Ortiz,
{\it A Note on the Semi-Classical Approximation
in Quantum Gravity}, gr-qc/9412040
\bibitem{jr} J. M. Jauch and F. Rorlich, The Theory of Photons and Electrons,
Springer Verlag, NY. (1980)
\bibitem{schwi} J. Schwinger, Phys. Rev. 82 (1951) 664
\bibitem{BD} N.D. Birrel and P.C.W. Davies, {\it Quantum Fields in Curved
Space},
Cambridge University Press (1982).
\bibitem{mtw} C. W. Misner, K. S. Thorne and J. A. Wheeler, {\it Gravitation}, 
Freeman, San Fransisco, (1973)
\bibitem{PKO} R. Parentani, J. Katz and  I.
Okamoto, Class. and Quant. Grav. {\bf 12} (1995) 1663
\bibitem{rec} R. Parentani, Nucl. Phys. B {\bf 454} (1995) 227  and 
Nucl. Phys. B {\bf 465} (1996) 175
\bibitem{wdwpt} R. Parentani, {\it Time dependent
perturbation theory in Quantum Cosmology}
LPTENS 96/46 preprint (1996) 
\bibitem{wdwpc} R. Parentani, {\it Pair creation in Quantum Cosmology},
in preparation

\end{thebibliography}
\end{document}